\documentclass[aip,jcp,preprint]{revtex4-1}
\usepackage{amssymb}
\usepackage{amsmath}
\usepackage{amsfonts}
\usepackage{graphicx}
\usepackage{tikz}
\usepackage{natbib}

\usepackage[size=small]{caption}
\usepackage{newfloat}
\DeclareFloatingEnvironment[
    fileext=loa,
    listname=List of Algorithms,
    name=ALGORITHM,
    placement=tbhp,
]{algorithm}

\DeclareCaptionFormat{algorithms}{\vskip-15pt\hrulefill\par#1#2#3\vskip-6pt\hrulefill}
\makeatletter
\renewcommand\@make@capt@title[2]{%
 \@ifx@empty\float@link{\@firstofone}{\expandafter\href\expandafter{\float@link}}%
  {\textbf{#1}}\@caption@fignum@sep#2\quad
}%
\makeatother

\begin{document}

\title{Nonlinear Intrinsic Variables and State Reconstruction in Multiscale Simulations}

\author{Carmeline J. Dsilva}
\email{cdsilva@princeton.edu}
\affiliation{Department of Chemical and Biological Engineering, Princeton University, Princeton, New Jersey, USA}

\author{Ronen Talmon}
\email{ronen.talmon@yale.edu}
\affiliation{Department of Mathematics, Yale University, New Haven, Connecticut, USA}

\author{Neta Rabin}
\email{netar@afeka.ac.il}
\affiliation{Department of Exact Sciences, Afeka Tel-Aviv Academic College of Engineering, Tel-Aviv, Israel}

\author{Ronald R. Coifman}
\email{coifman@math.yale.edu}
\affiliation{Department of Mathematics, Yale University, New Haven, Connecticut, USA}

\author{Ioannis G. Kevrekidis}
\email{yannis@princeton.edu}
\affiliation{Department of Chemical and Biological Engineering, Princeton University, Princeton, New Jersey, USA}
\affiliation{Program in Applied and Computational Mathematics, Princeton University, Princeton, New Jersey, USA}

\date{\today}

\begin{abstract}
Finding informative low-dimensional descriptions of high-dimensional simulation data
(like the ones arising in molecular dynamics or kinetic Monte Carlo simulations of
physical and chemical processes) is crucial to understanding physical phenomena, and can
also dramatically assist in accelerating the simulations themselves.
In this paper, we discuss and illustrate the use of nonlinear intrinsic variables (NIV)
in the mining of high-dimensional multiscale simulation data.
In particular, we focus on the way NIV allows us to functionally merge different
simulation ensembles, and {\em different partial observations of these ensembles}, as well
as to infer variables not explicitly measured.
The approach relies on certain simple features of the underlying process variability to
filter out measurement noise and systematically recover a unique reference coordinate frame.
We illustrate the approach through two distinct sets of atomistic simulations:
a stochastic simulation of an enzyme reaction network exhibiting both fast and slow time scales,
 and a molecular dynamics simulation of alanine dipeptide in explicit water.

\end{abstract}


\maketitle

\section{Introduction}
The last decade has witnessed extensive advances in dimensionality reduction techniques:
finding meaningful low-dimensional descriptions of high-dimensional data \cite{Tenenbaum2000,Roweis2000,Donoho2003,Belkin2003,Coifman2006}.
These developments have the potential to significantly enable the computational exploration
of physicochemical problems.
If the (high-dimensional) data $\mathbf{Y}(t)$ arise from, for example, a
molecular dynamics simulation of a macromolecule in solution, or from the stochastic
simulation of a complex chemical reaction scheme, the detection of a few good, coarse-grained
``reduction coordinates" $\mathbf{x}(t)$ can be invaluable in understanding and predicting system behavior.

While the benefits from such reduced descriptions are manifest, a crucial shortcoming of
 data-driven reduction coordinates is their dependence on the specific data set processed,
and not only on the physical model in question.
It is well known that, even in the simple linear case of Principal Component Analysis \cite{jolliffe2005principal},
different data sets on the same low-dimensional hyperplane in the ambient space
will lead to different basis vectors - in effect, to different reduction coordinates $\mathbf{x}$.
While this can be easily rectified by an affine transformation
(see, by analogy, the discussion in Lafon {\em et al.}\cite{lafon2006data}),
the problem becomes
exacerbated when the low-dimensional space is curved (a manifold, rather than a hyperplane)
and when different data sets are obtained using different instrumental modalities
(such as when one wants to merge molecular dynamics data with, for example, spectral information).
Clearly, the ability to systematically construct a {\em unique} and {\em consistent} reduction
coordinate set, shared by all measurement ensembles and observation modalities, is invaluable.
We will call these coordinates Nonlinear Intrinsic Variables.
Embedding data in such a coordinate system allows us to naturally merge different observations of the same system;
more importantly, it enables the construction of an empirical mapping between these different
observation ensembles, allowing us to complete partial measurements in a test data set from a training data set
that consists of {\em different} observations.
To construct this empirical mapping and the associated observers,
accurate interpolation tools must be available in the embedding space; to this
end, we will demonstrate the use of a multiscale Laplacian Pyramid approach \cite{rabin2012heterogeneous}.

We will illustrate our methodologies with two distinct examples.
The first is a simulation
of two Goldbeter-Koshland modules in an enzyme kinetics model using the Gillespie Stochastic Simulation
Algorithm (SSA) \cite{gillespie1977exact};
in certain parameter regimes, separation of time scales is known
to reduce the ODE model of this kinetic scheme to an effective two-dimensional description \cite{zagaris2012stability}.
Although this example is rather simple, it will serve as an introduction to our techniques and highlight the main features of the algorithms.
The second example is a molecular dynamics
simulation (in explicit water) of a simple peptide fragment (alanine dipeptide) whose folding
dynamics are known to be described through a small set of physical observables \cite{bolhuis2000reaction}.
This example will allow us to compare our approach to more common techniques,
such as diffusion maps \cite{coifman2005geometric} for dimensionality reduction
and nearest neighbor interpolation for observation reconstruction.
The remainder of the paper is structured as follows: in Section \ref{sec:NIV} we present the Nonlinear Intrinsic Variable formulation and
the associated inference method.
Section \ref{sec:LapPyr} contains our discussion of Laplacian Pyramids that
is used for the completion of partial observations.
In Section \ref{sec:examples}, the results
of the application of the approach to simulation data from our two illustrative examples are presented and discussed.
We conclude with a summary and our perspective on open issues in Section \ref{sec:conclusions}.

\section{Nonlinear Intrinsic Variables} \label{sec:NIV}

\subsection{Overview}
Let $\mathbf{Y}(t)$ be a high-dimensional measured process in $\mathbb{R}^n$ consisting of $n$ observable variables.
We impose two critical assumptions. First, the measured process is assumed to be a manifestation in an observable domain of a low-dimensional diffusion process. Thus, it can be expressed by
\begin{equation}
	\mathbf{Y}(t) = f(\mathbf{x}(t)),
\end{equation}
where $f:\mathbb{R}^d \rightarrow \mathcal{M}$ is an unknown (possibly nonlinear) function, $\mathcal{M} \subset \mathbb{R}^n$ is a $d$-dimensional manifold,
and $\mathbf{x}(t)$ is a diffusion process that consists of $d$ underlying variables (with $d \ll n$).
Second, the dynamics of the diffusion process in each of its underlying variables are described by normalized stochastic differential equations as
\begin{equation}
	d x_i(t) = a_i (x_i(t)) dt + d w_i(t), \ i=1,\ldots,d,
\end{equation}
where $a_i$ are unknown drift functions and $\dot{w}_i(t)$ are independent white noises.
The independence is our second critical assumption.

\begin{figure*}[ht]
\includegraphics[width=5in]{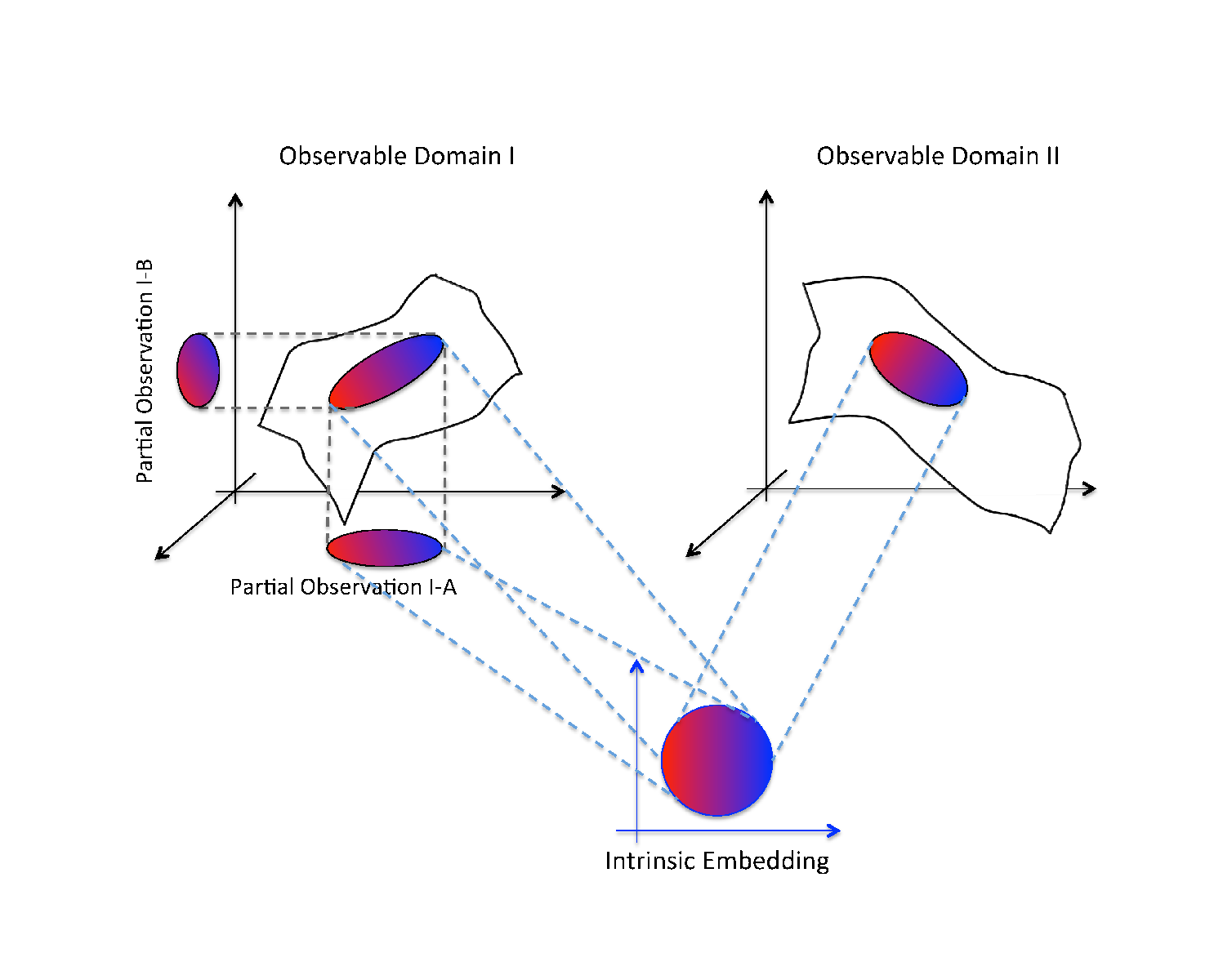}
\caption{Illustration of the nonlinear embedding that yields an intrinsic representation independent of the measurement function $\mathbf{f}$. (Bottom) The underlying variables in which the noises are independent with unit variance. The circle illustrates samples from, say, a short trajectory in time that sample a disc on the manifold. (Top Left) The first set of observed variables. The ellipse illustrates the mapping of the sphere of the underlying samples into the observable domain via the first observation function. In this sketch, we illustrate that the observations might be {\em partial}, i.e., might consist of merely a subset of the observed domain variables. (Top Right) Second set of observable variables. The ellipse illustrates the mapping of the sphere of the underlying samples into this (different) observable domain via a second observation function.
\label{fig:IntrinsicIllustration}}
\end{figure*}

Given a sequence of samples $\mathbf{Y}(t), \ t=1,\ldots,T$, we present an empirical method to construct a unique and consistent reduction coordinate set, represented here by $\mathbf{x}(t)$ \cite{singer2008non}.
Because the empirical method we will describe is independent of the observation function $\mathbf{f}$,
we refer to the coordinates of $\mathbf{x}(t)$ as Nonlinear Intrinsic Variables (NIV).
The available samples $\mathbf{Y}(t)$ may be the result of different measurement functions $\mathbf{f}$ in various observable domains,
or they may be partial measurements consisting of merely a subset of the coordinates of the observable domains.
The idea is to empirically construct a NIV coordinate system driven entirely by measurements that is invariant to the observation function $\mathbf{f}$
(see Figure \ref{fig:IntrinsicIllustration} for a schematic illustration).
We remark that the available data should be ``rich enough'', i.e., consist of a sufficient amount of historical data with adequate variability, in order to obtain the full empirical model.

The method consists of the following main principles.
(1) The underlying diffusion process implies that a short trajectory of successive samples mainly consists of diffusion noise,
and hence, creates a ``sphere" of samples in the underlying domain $\mathbb{R}^d$.
%
This sphere is mapped to an ellipse in the observable domain by the measurement function $\mathbf{f}$.
In this work, the identification of the associated ellipse of samples according to the time trajectory of our data enables us
to estimate the tangent planes of the observable manifolds {\em via the principal components of the covariance matrices of the samples in these ellipses}.
(2) The principal directions of the tangent planes are utilized to define a Riemannian metric that is shown to be {\em locally invariant to the measurement function $\mathbf{f}$}.
(3) The NIV are constructed through the eigenvalue decomposition of a Laplace operator that is built upon a pairwise affinity between the samples, defined using this Riemannian metric.

\subsection{Mahalanobis Distance}
\label{subsec:mahalanobis}

Let $\mathbf{C}(t)$ be the covariance matrix associated with the measured sample $\mathbf{Y}(t)$. In practice, the covariance matrix can be estimated from a short trajectory of samples in time around the sample $\mathbf{Y}(t)$ by
\begin{equation}
	\widehat{\mathbf{C}}(t) = \sum \limits _{\tau = t-L}^{t+L} (\mathbf{Y}(\tau) - \widehat{\boldsymbol{\mu}}(t))(\mathbf{Y}(\tau) - \widehat{\boldsymbol{\mu}}(t))^T,
	\label{eq:cov}
\end{equation}
where $\widehat{\boldsymbol{\mu}}(t)$ is the empirical mean of the short trajectory of samples.
We define a Riemannian metric between a pair of samples using the associated covariance matrices as
\begin{equation}
	d^2(\mathbf{Y}(t), \mathbf{Y}(\tau)) = 2 (\mathbf{Y}(t) - \mathbf{Y}(\tau))^T(\widehat{\mathbf{C}}(t) + \widehat{\mathbf{C}}(\tau))^{\dagger}(\mathbf{Y}(t) - \mathbf{Y}(\tau));
	\label{eq:mahalanobis}
\end{equation}
this is the Mahalanobis distance (and $^{\dagger}$ denotes a pseudoinverse, as discussed below).
As previously described, the covariance matrices convey the local variability of the measurements and are utilized to explore and learn the tangent planes of the observable manifold.
This information is then utilized in \eqref{eq:mahalanobis} to compare a pair of points according to the directions of their respective tangent planes.
The Mahalanobis distance is invariant under affine transformations.
Thus, by assuming that the observation function $\mathbf{f}$ is bi-Lipschitz and smooth, and by using local linearization of the function, i.e., $\mathbf{Y}(t) = \mathbf{J}(t) \mathbf{x}(t) + \boldsymbol{\epsilon}(t)$ where $\mathbf{J}(t)$ is the Jacobian of $\mathbf{f}(\mathbf{x}(t))$ and $\boldsymbol{\epsilon}(t)$ is the residual consisting of higher-order terms, it was shown by Singer and Coifman \cite{singer2008non} that $\mathbf{C}(t) = \mathbf{J}(t)\mathbf{J}^T(t)$ and that the Mahalanobis distance approximates the Euclidean distance between the corresponding samples of the underlying process to second order, i.e.,
\begin{equation}
	\| \mathbf{x}(t) - \mathbf{x}(\tau) \|^2 = d^2(\mathbf{Y}(t), \mathbf{Y}(\tau)) + \mathcal{O}(\| \mathbf{Y}(t) - \mathbf{Y}(\tau)\|^4).
\end{equation}
This result implies that the Mahalanobis distance is invariant to the measurement function $\mathbf{f}$, and hence,
it yields the same distances between samples obtained under different observation functions or even partial observations.
We would like to note that, in general, $\mathbf{f}$ being bi-Lipschitz implies that $\mathbf{f}$ is invertible (on the $d$-dimensional
manifold $\mathcal{M}$).
However, in practice, determining whether $\mathbf{f}$ contains sufficient information and is ``rich enough'' to completely determine the underlying process is a non-trivial task.
In this work, we exploit the fact that $\mathbf{C}(t) = \mathbf{J}(t)\mathbf{J}^T(t)$, which implies that $\mathbf{C}(t)$ is an $n \times n$ positive semidefinite matrix of rank $d$, to empirically infer the dimension $d$.
According to the spectrum of the local covariance matrices and their corresponding spectral gaps, we approximate the rank of the matrices.
Consistent rank estimates among these local covariance matrices are taken to imply that the measurements are ``rich enough", and hence, may be good indicators for the dimension $d$.
Since the dimension $d$ of the underlying process is typically considerably smaller than the dimension of the measured process $n$,
the covariance matrix is singular and non-invertible;
thus, we use the pseudo-inverse in \eqref{eq:mahalanobis}.

\subsection{Laplace Operator}
The Mahalanobis distance described in Section \ref{subsec:mahalanobis} enables us to compare observations in terms of the intrinsic variables of the associated underlying diffusion process.
In this section, we show how to recover the underlying process itself from the pairwise Euclidean distances through the eigenvectors of a Laplace operator.

Let $\mathbf{W}$ be a pairwise affinity matrix (kernel) based on a Gaussian, whose $(t,\tau)$-th element is given by
\begin{equation}
	W_{t,\tau} = \exp \left\{ - \frac{ d^2(\mathbf{Y}(t), \mathbf{Y}(\tau) )} {\varepsilon}\right\},
	\label{eq:kernel}
\end{equation}
where $\varepsilon$ is the kernel scale, which can be set according to Hein and Audibert \cite{hein2005intrinsic} and Coifman {\em et al.} \cite{coifman2008graph}.
Based on the kernel, we form a weighted graph, where the measurements $\mathbf{Y}(t)$ are the graph nodes and the weight of the edge connecting node $\mathbf{Y}(t)$ to node $\mathbf{Y}(\tau)$ is $W_{t,\tau}$.
In particular, a Gaussian kernel exhibits a notion of locality by defining a neighborhood around each measurement $\mathbf{Y}(t)$ of radius $\varepsilon$,
i.e., measurements $\mathbf{Y}(\tau)$ such that $d^2(\mathbf{Y}(t), \mathbf{Y}(\tau) ) > \varepsilon$ are weakly connected to $\mathbf{Y}(t)$.
In practice, we set $\varepsilon$ to be the median of the pairwise distances.
According to the graph interpretation, this implies a well-connected graph because each measurement is effectively connected to half of the other measurements \cite{rohrdanz2011determination}.

Let $\mathbf{D}$ be a diagonal matrix whose elements are the row sums of $\mathbf{W}$, and let $\mathbf{W}^{\mathrm{norm}} = \mathbf{D}^{-1/2}\mathbf{W}\mathbf{D}^{-1/2}$
be a normalized kernel that shares its eigenvectors with the normalized graph-Laplacian $\mathbf{I}-\mathbf{W}^{\mathrm{norm}} \:$ \cite{chung1997spectral}.
The eigenvectors of $\mathbf{W}^{\mathrm{norm}}$, denoted $\psi_j$, reveal the underlying structure of the data \cite{coifman2005geometric}.
Specifically, the $i$-th coordinate of the $j$-th eigenvector can be associated with an intrinsic coordinate $j$ of the sample $\mathbf{x}(i)$ of the underlying process.
The eigenvectors are ordered such that $|\lambda_1| \ge |\lambda_2| \ge \dots \ge |\lambda_n|$, where $\lambda_j$ is the eigenvalue associated with eigenvector $\psi_j$.
Because $\mathbf{W}^{\mathrm{norm}} \sim \mathbf{D}^{-1}\mathbf{W} $, and $\mathbf{D}^{-1}\mathbf{W}$ is row-stochastic,
$\lambda_1 = 1$ and $\psi_1$ is the diagonal of $\mathbf{D}^{1/2}$.
The next few eigenvectors can be argued to describe the geometry of the underlying manifold \cite{coifman2005geometric}.
However, some eigenvectors can be higher harmonics of the same principal direction along the data manifold.
This is analogous to how the eigenfunctions $\cos x$ and $\cos 2x$ of the usual Laplacian in one spatial dimension
and with no flux boundary conditions are one-to-one with the values of $x$ for $0 \le x \le 1$;
one must check for correlations between the eigenvectors before selecting those that describe the underlying manifold geometry.
The above steps to construct the nonlinear intrinsic variables are summarized in Algorithm \ref{algo}.

Ignoring the higher harmonics, each retained eigenvector then describes an intrinsic variable for the data set of interest.
%
We must normalize the eigenvectors from different data sets so that the resulting embeddings are consistent.
We first scale the eigenvectors so that $\|\psi_i\| = T$, where $T$ is the number of data points,
to make the embedding coordinates invariant to the size of the data set.
%
Still, the computed embedding eigenvectors, even for two identical data sets, may differ by a sign.
Reconciling the signs for the embeddings of different data sets can be rationally done in several ways and is somewhat problem-specific.
For example, if the mean of the embedding is sufficiently far from 0, we can require $\langle \psi_i \rangle > 0$;
alternatively, if there is a common region sampled by both data sets, the sign of each eigenvector can be chosen to optimize the consistency of the embeddings of the common region data.
We will return to the issue of embedding consistency for different data sets in our concluding discussion; for the moment,
we will assume that our different sets sample the same region of data space in a representative enough way such that the
correspondence between the sequences of retained eigenvectors for different embeddings is obvious.

\begin{algorithm}[th!]
\caption{Nonlinear Intrinsic Variables Construction}
\begin{enumerate}

\item
Obtain a sequence of high-dimensional observation samples $\mathbf{Y}(t)$.

\item
Compute the empirical covariance matrix $\widehat{\mathbf{C}}(t)$ of each sample $\mathbf{Y}(t)$ {\em in a short window in time} according to \eqref{eq:cov}.

\item
Using the samples and their associated covariance matrices, compute the Mahalanobis distance between the observations \eqref{eq:mahalanobis} .

\item
Build the pairwise affinity matrix $\mathbf{W}$ and the corresponding normalized kernel $\mathbf{W}^{\mathrm{norm}}$ (\ref{eq:kernel}).

\item
Apply eigenvalue decomposition to the normalized kernel and view the values of its principal eigenvectors (modulo the possibility of
 ``higher harmonics", see text) as the Nonlinear Intrinsic Variables (NIV) of the given observations.

\end{enumerate}
\hrule
\label{algo}
\end{algorithm}

\section{Laplacian Pyramids for Data Extension} \label{sec:LapPyr}

In this work, we are not only interested in extracting the underlying variables $\mathbf{x}$ from some (partial) observations $\mathbf{Y}$,
but also interested in extending high-dimensional functions on a set of points which lie in a low-dimensional space.
More specifically, viewing the ambient space coordinates $\mathbf{Y}$ as functions on the low-dimensional data $\mathbf{x} \in \mathbb{R}^d$,
we want to estimate $\mathbf{Y}$ for new points $\mathbf{x}$.
Laplacian Pyramids (LP) is a multiscale algorithm for extending an empirical function $\mathbf{f}$ defined on a set of points
to new points not in the dataset.
The algorithm uses Laplacian kernels of decreasing widths to create multiscale representations of $\mathbf{f}$;
these representations can be easily extended to new data points.
This type of multiscale representation was introduced by Burt and Adelson \cite{burt1983laplacian} for image coding,
and was later shown to be a tight frame by Do and Veterli \cite{do2003framing}.
Recently, LP was used to extend nonlinear embedding coordinates to new high-dimensional data points \cite{rabin2012heterogeneous}.
We will first review the LP algorithm for approximating and extending a one-dimensional function,
and then describe the application of LP in extending high-dimensional functions.

Let $\mathbf{f}: \Gamma \rightarrow \mathbb{R}^n$ be a function that is known on a subset of points $S \subset \Gamma$.
A coarse representation of $\mathbf{f}$ is generated using a coarse smoothing operator $P_0$.
The smoothing operator $P_0$ is a normalized, coarse Laplacian kernel, defined by
\begin{equation}
p_0(i, j)= s_0^{-1}(i)w_0(i, j),\: i, j \in \Gamma,
\end{equation}
where $w_0(i, j)=e^{-d^2(i, j) / \sigma_0}$ and $s_0(i)=\sum_{j \in S}w_0(i, j)$ is the normalizing term.
The pairwise distance $d(i, j)$ is typically the Euclidean distance, and the parameter $\sigma_0$ is set to be large compared to the values of $d^2(i, j)$.
The application of $P_0$ to $\mathbf{f}$ yields a coarse representation of the function, which we denote by $\mathbf{f}_0=P_0(\mathbf{f})$.

The difference $\delta_1 = \mathbf{f}-P_0(\mathbf{f})$ is the input for the next iteration of the algorithm,
which uses the smoothing operator $P_1$, $P_1 \propto e^{-d^2(i, j)^2 / 2^{-1} \sigma_0}$, to construct a coarse representation of $\delta_1$.
The obtained representation of $\delta_1$, $P_1(\delta_1)$ together with the result of the previous iteration $\mathbf{f}_0 = P_0(\mathbf{f})$
yields a new, finer representation of $\mathbf{f}$, $\mathbf{f}_1 = P_0(\mathbf{f}) + P_1(\delta_1)$.
In an iterative manner, multiscale representations of the function $\mathbf{f}$, denoted $\mathbf{f}_l$, are constructed.
\begin{equation} \label{eq:LP_multi_scale}
 \begin{array}{cl}
\mbox{scale 0:} & \mathbf{f}_0 = P_0(\mathbf{f}) \\
\mbox{scale 1:} & \mathbf{f}_1 = P_0(\mathbf{f}) + P_1(\delta_1) \\
: & : \\
\mbox{scale l:} & \mathbf{f}_l = P_0(\mathbf{f}) + \sum_{k=1}^{l}P_k(\delta_k)\\
\end{array}
\end{equation}
As $l$ increases, the approximation becomes more refined because $P_l \propto e^{-d^2(i, j) / 2^{-l} \sigma_0}$ uses a Laplacian kernel of a finer width.
The iterations stop when the difference between $\mathbf{f}$ and $\mathbf{f}_l$ is smaller than a pre-defined error threshold.

The representations $\mathbf{f}_0, \mathbf{f}_1, \dots, \mathbf{f}_l$ can be extended to a new point $\tilde{i} \in \Gamma \backslash S $ by extending the operators $P_0, P_1,\ldots,P_l$.
For example, $\mathbf{f}_0(\tilde{i}) = \sum_{i \in S} p_0(\tilde{i}, i)\mathbf{f}(i)$ and
$\mathbf{f}_1(\tilde{i}) = \mathbf{f}_0(\tilde{i}) + \sum_{(i \in S)}p_1(\tilde{i}, i)\delta_1(i)$.

\begin{figure*}[ht]
\includegraphics[width=6in]{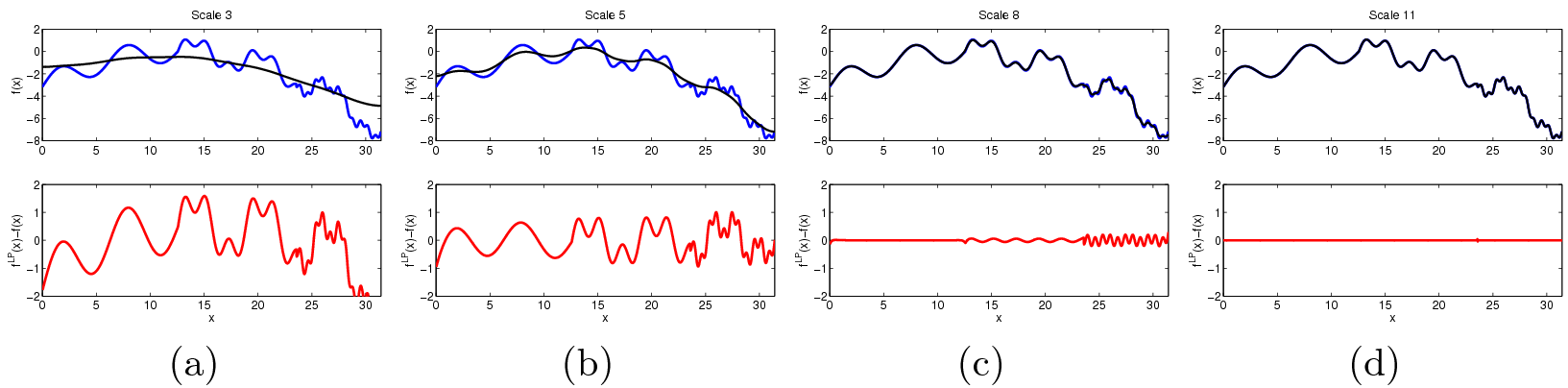}
\caption{Approximation of the function $\mathbf{f}(x)$ defined in \eqref{eq:LP_example} using Laplacian Pyramids for ``scales'' (a) 3, (b) 5, (c) 8, and (d) 11. The top shows the true function in blue and the LP approximation is shown in black, and the bottom shows the residual error in the LP approximation.}
\label{fig:LP_ex}
\end{figure*}

\begin{widetext}
Figure \ref{fig:LP_ex} displays an illustrative example of the algorithm, when applied to the function
 \begin{equation} \label{eq:LP_example}
\mathbf{f}(x) = \left\{
\begin{array}{l l}
-0.02(x-4\pi)^2 + \sin(x) &  0 \le x \le 4\pi \\
-0.02(x-4\pi)^2 + \sin(x) + 0.5 \sin(3x) &  4\pi < x \le 7.5 \pi \\
-0.02(x-4\pi)^2 + \sin(x) + 0.5 \sin(3x) +0.25 \sin(9x) &  7.5 \pi < x \le 10 \pi
\end{array}
\right.
\end{equation}
that contains several scales.
The coarse regions of the function ($0 \le x \le 4\pi$) are well approximated by a small number of scales.
As the function becomes more oscillatory ($\pi \le x \le 7.5\pi$ and $7.5\pi \le x \le 10\pi$),
a finer representation, and a larger number of scales  $l$, is required to capture its behavior.
\end{widetext}

In this work, LP is applied to extend a high-dimensional function $\mathbf{f}:\mathbb{R}^d \rightarrow \mathcal{M}$, 
which maps a set of points in the NIV space to their values $\mathbf{Y}(i)$ in the observable space.
Let $\Psi(i) = \left(\psi_1(i),\psi_2(i),\ldots,\psi_d(i)\right)$ be the set of NIV that were constructed from the data samples $\mathbf{Y}(i)$, as described in Algorithm \ref{algo}.
The values of function $\mathbf{f}:\mathbb{R}^d \rightarrow \mathcal{M}$ are known on the subset $S = \{\Psi(i)\}$, with $\mathbf{f}(\Psi(i)) = \mathbf{Y}(i)$.

A na\"{\i}ve way to extend $\mathbf{f}$ to a new data point $\Psi(\tilde{i})$ is find the point's nearest neighbors in NIV space and average their function values.
A different, point-wise adaptive approach is described by Buchman {\em et al.} \cite{buchman2011high}:
high-dimensional hurricane tracks were estimated from low dimensional embedding coordinates using a weighted average of the points close to $\Psi(\tilde{i})$ in the embedded space.
However, this point-wise adaptation requires setting the nearest neighborhood radius parameter for every point.
The LP algorithm finds the appropriate nearest neighborhood radius for each new point $\Psi(\tilde{i})$.
This radius will be large in smooth regions of the function, and small in regions in which $\mathbf{f}$ contains higher frequency components.
The LP approximation of a new, high dimensional point $\mathbf{Y}(\tilde{i})$ is calculated by a weighted average of the function values that belong to the neighboring points.
The weights are based on the pairwise distances in the intrinsic, low-dimensional space.
In practice, a set of smoothing operators $P_0, P_1, \ldots, P_l$, with
\begin{equation} \label{eq:LP_multi_scale_app}
P_l \propto e^{-d^2(\Psi(i),\Psi(j)) / 2^{-l} \sigma_0},
\end{equation}
are constructed and later extended to create the multiscale approximations as defined in Eq. \ref{eq:LP_multi_scale}.
The LP algorithm for the inverse mapping is summarized in Algorithm \ref{algo_LP}.

\begin{algorithm}[th!]
\caption{Laplacian Pyramids for Inverse Mapping}
\begin{enumerate}

\item
Construct a set of smoothing operators $P_0, P_1, \ldots, P_l$  based on intrinsic pairwise distances $d(\Psi(i),\Psi(j))$ (where $d$ is typically the Euclidean distance).

\item
Use the smoothing operators to obtain a multiscale representation (see Eq. \ref{eq:LP_multi_scale_app}) of $\mathbf{f}:\Psi(i) \rightarrow Y(i)$.

\item Given a new point $\Psi(\tilde{i})$ in NIV, extend the smoothing operators $P_0, P_1, \ldots, P_l$  by
$p_l(\Psi(\tilde{i}), \Psi(i)) = s_l^{-1}(\Psi(i)) w_l(\Psi(\tilde{i}),\Psi(i))$.

\item
Use the extended smoothing operators to approximate the value of $Y(\tilde{i})$ as $Y(\tilde{i}) \approx \mathbf{f}_l(\Psi(\tilde{i})) = \mathbf{f}_0(\Psi(\tilde{i})) + \sum_{k}\sum_{(i \in S)}p_k(\Psi(\tilde{i}), \Psi(i))\delta_k(\Psi(i))$


\end{enumerate}
\hrule
\label{algo_LP}
\end{algorithm}

\section{Models and Results} \label{sec:examples}

\subsection{A Chemical Reaction Network}\label{subsec:rxn_network}

We first consider a chemical reaction network involving  multiple enzyme-substrate interactions \cite{zagaris2012stability}.
The reaction steps that comprise the network are\\
\begin{equation}
\begin{array}{rcl}
E + S \overset{e_1}{\underset{e_{-1}}{\leftrightharpoons}} & E:S & \overset{e_2}{\rightarrow} E + S^{*} \\
S + E \overset{b_1}{\underset{b_{-1}}{\leftrightharpoons}} & S:E & \overset{b_2}{\rightarrow} S + E^{*}\\
D + S^{*} \overset{d_1}{\underset{d_{-1}}{\leftrightharpoons}} & D:S^{*} & \overset{d_2}{\rightarrow} D + S\\
F + E^{*} \overset{f_1}{\underset{f_{-1}}{\leftrightharpoons}} & F:E^{*} & \overset{f_2}{\rightarrow} F + E
\end{array}
\end{equation}

The ``$^{*}$'' denotes an activated form of a species, and the ``:'' denotes a complex formed between two species; the complexes $E:S$ and $S:E$ are not equivalent.
There are 10 species in this reaction system.
However, one can write four conservation equations (since total $E$, $S$, $D$, and $F$ are all conserved) to reduce the system to 6 dimensions
(which we order as $S$, $E$, $E:S$, $S:E$, $D:S^{*}$, $F:E^{*}$).
We consider a parameter regime in which the ODE approximation of this scheme exhibits a separation of time scales, so that initial conditions quickly approach a two-dimensional manifold.
Details about the specific parameter values can be found in Appendix \ref{app:rxn}.

Although the dynamics of chemical reaction networks are typically described by a system of ODEs, the ODEs are only an approximation that holds
in the limit of a large number of molecules.
When the number of molecules is small, the system is inherently stochastic and its dynamics can be simulated using the
Gillespie Stochastic Simulation Algorithm (SSA) \cite{gillespie1977exact}; at intermediate molecule counts, the chemical Langevin approximation \cite{gillespie2000chemical}
becomes useful.
We can control the level of noise in our simulation by adjusting the volume $V$, and therefore, adjusting the number of molecules, in the system.
We take the volume small enough so that we can still observe appreciable stochasticity in small simulation bursts, but large enough (in our simulations, we take $V=10^5$) so that the underlying two-dimensional manifold is (relatively) smooth.

\begin{figure*}[ht]
  \includegraphics[width=6in]{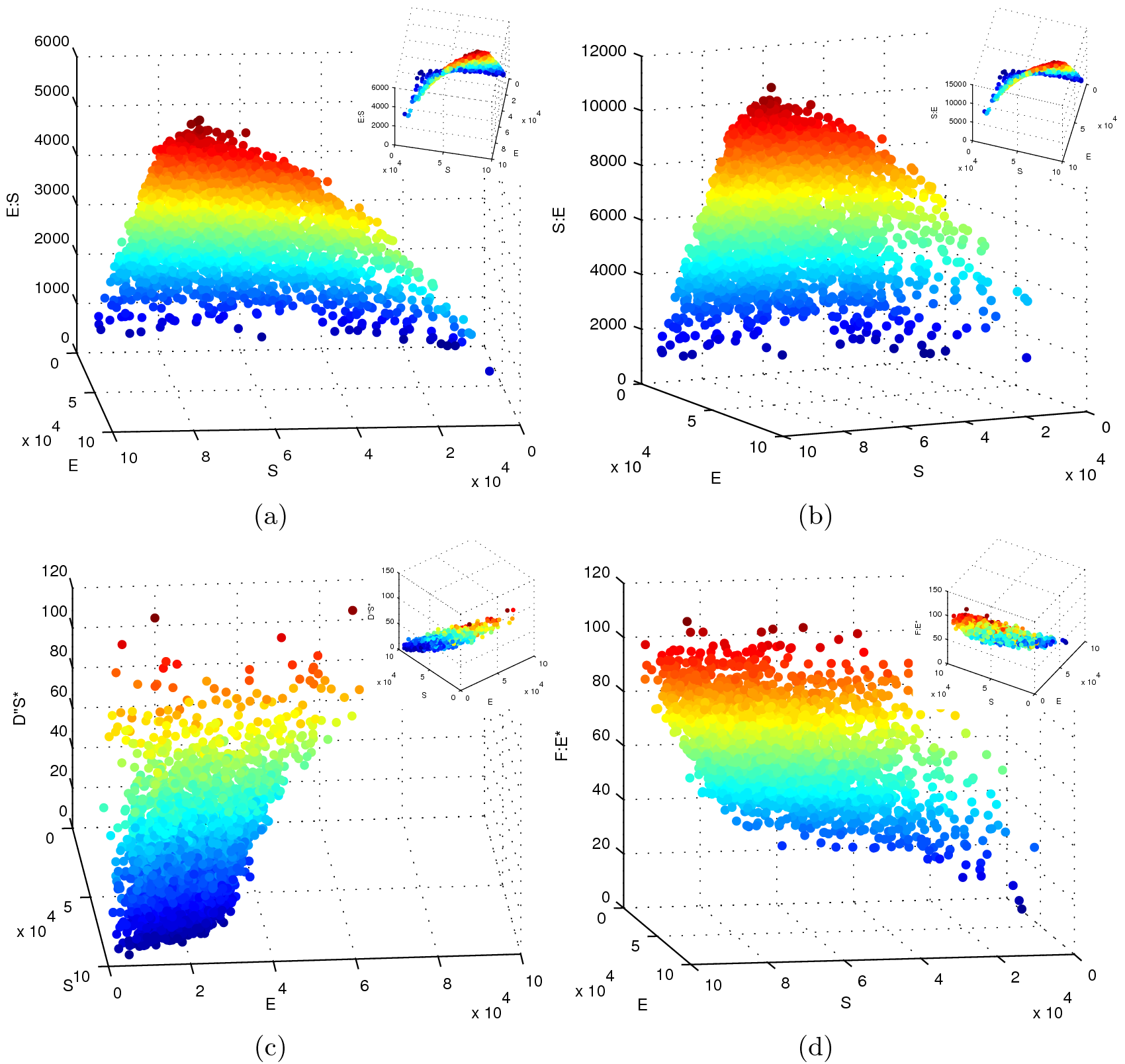}
    \caption{Projections of the data obtained from stochastic simulation of the chemical reaction network described in Section \ref{subsec:rxn_network}. The insets show rotations of the projections to illustrate the approximate two-dimensionality of the ``slow manifold''.}
    \label{fig:rxn_manifolds}
\end{figure*}

We generate 3000 random initial conditions $\mathbf{Y}_0(1), \dots, \mathbf{Y}_0(3000) \in \mathbb{R}^6$, enforcing that all concentrations
must be non-negative.
We evolve each point $\mathbf{Y}_0(t)$ forward for 10 time units using the SSA to obtain a point $\mathbf{Y}(t) \in \mathbb{R}^6$;
according to the time scales calculated from the linearized ODEs, 10 time units is sufficiently long for the initial points in the ODE system to converge to the two-dimensional manifold,
but not long enough for the points to converge to a one-dimensional curve or to the final steady state (see Appendix \ref{app:rxn} for more details).
In our stochastic simulations, the initial points appear to converge to an approximate two-dimensional manifold
(in expected value, see Figure \ref{fig:rxn_manifolds}).
We consider $\mathcal{Y} = \{ \mathbf{Y}(t): t=1, \dots, 3000 \}$ to be representative points ``on" this apparent two-dimensional manifold.
From each manifold point $\mathbf{y} \in \mathcal{Y}$, we run 20 short simulation ``bursts'', each for 0.2 time units.
We denote the endpoints from the short simulations as $\mathcal{Y}^{burst}(\mathbf{y})$.

We consider two different data sets from our simulations.
Data set 1, denoted $\mathcal{Y}_1$, consists of $\mathbf{Y}(1), \dots, \mathbf{Y}(2000)$, restricted to components $S$, $E$, $S:E$, and $F:E^{*}$, i.e.,
$$\mathcal{Y}_1 = \left\{
\left( \begin{array}{cccccc}
1 & 0 & 0 & 0 & 0 & 0 \\
0 & 1 & 0 & 0 & 0 & 0 \\
0 & 0 & 0 & 1 & 0 & 0 \\
0 & 0 & 0 & 0 & 0 & 1
\end{array} \right) \mathbf{Y}(t) \in \mathbb{R}^4: t=1, \dots, 2000 \right\}.$$
Data set 2, denoted $\mathcal{Y}_2$, consists of $\mathbf{Y}(1500), \dots, \mathbf{Y}(3000)$, restricted to components $S$, $E$, $E:S$, and $D:S^{*}$, i.e.,
$$\mathcal{Y}_2 = \left\{
\left( \begin{array}{cccccc}
1 & 0 & 0 & 0 & 0 & 0 \\
0 & 1 & 0 & 0 & 0 & 0 \\
0 & 0 & 1 & 0 & 0 & 0 \\
0 & 0 & 0 & 0 & 1 & 0
\end{array} \right)
\mathbf{Y}(t) \in \mathbb{R}^4: t=1500, \dots, 3000 \right\}.$$
The endpoints of the simulation bursts for the two data sets, $\mathcal{Y}^{burst}_1(\mathbf{y})$ and $\mathcal{Y}^{burst}_2(\mathbf{y})$, are defined analogously.
We then estimate the covariances for each point in each data set as
\begin{equation}
\widehat{\mathbf{C}}_i(\mathbf{y}) = \sum_{\mathbf{z} \in \mathcal{Y}^{burst}_i(\mathbf{y})} \left( \mathbf{z} - \hat{\mathbf{\mu}}_i(\mathbf{y}) \right)\left( \mathbf{z} - \hat{\mathbf{\mu}}_i(\mathbf{y}) \right)^T, \mathbf{y} \in \mathcal{Y}_i, i \in \{1, 2\}
\end{equation}
where $\hat{\mathbf{\mu}}_i(\mathbf{y})$ is the empirical mean of $\mathcal{Y}^{burst}_i(\mathbf{y})$.

\begin{figure*}[ht]
    \includegraphics[width=6in]{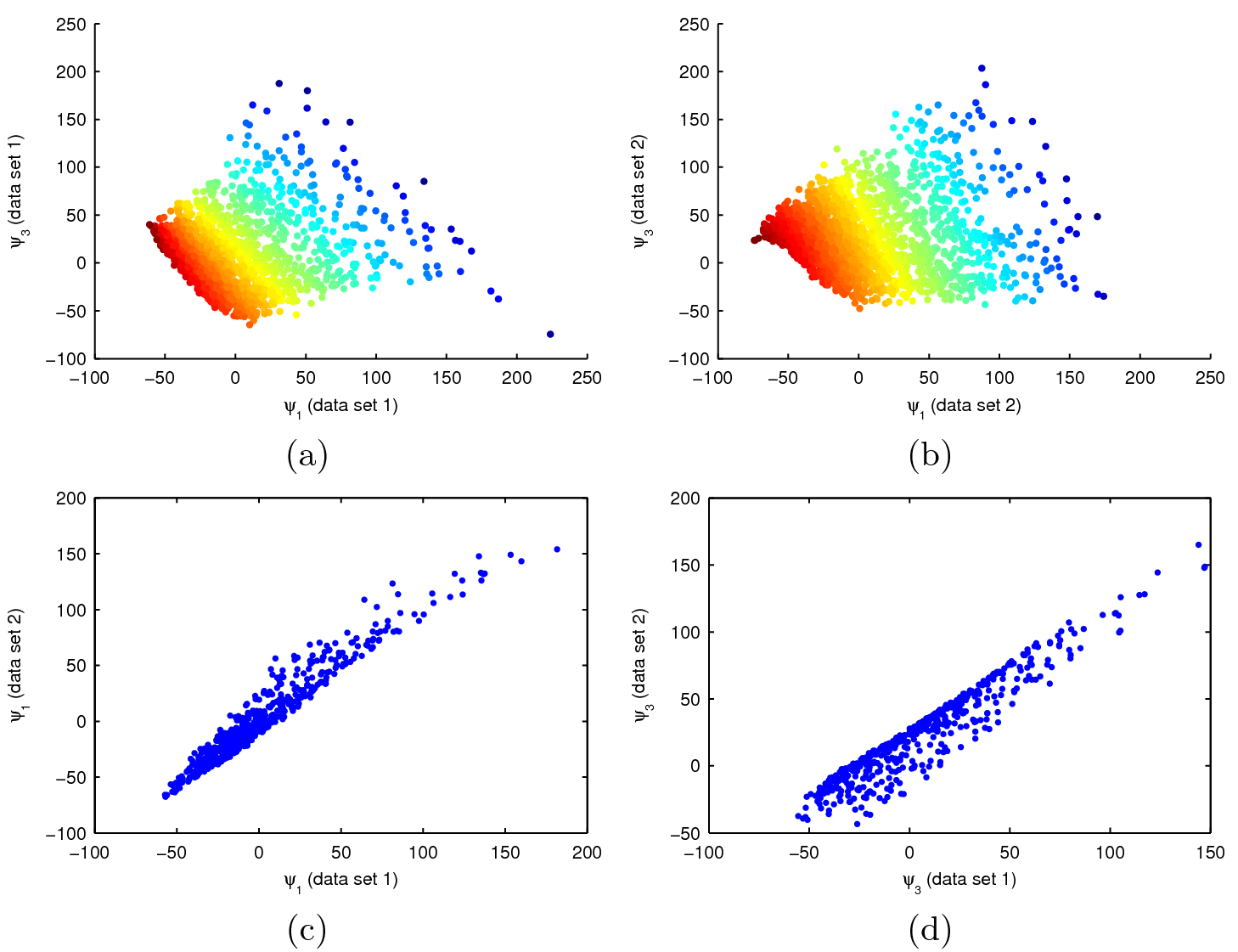}
    \caption{(a) NIV embedding obtained from $\mathcal{Y}_1$ (observations of components E, S, S:E, and F:E$^{*}$), colored by $S$. (b) NIV embedding obtained from $\mathcal{Y}_2$ (observations of components E, S, E:S, and D:S$^{*}$), colored by $S$. Visually, we can see that the embeddings obtained from $\mathcal{Y}_1$ and $\mathcal{Y}_2$ are consistent, even though the two data sets consist of observations of different chemical species. (c) Correlation of first NIV between two different embeddings (correlation=0.97). (d)  Correlation of second NIV between two different embeddings (correlation=0.95). We obtain a good quantitative agreement between the embedding coordinates for the two data sets.}
    \label{fig:rxn_embedding}
\end{figure*}

We first demonstrate that NIV produces the same embeddings for $\mathcal{Y}_1$ and $\mathcal{Y}_2$, even though the two data sets contain information of different chemical species.
Figure \ref{fig:rxn_embedding} shows the two-dimensional NIV embeddings for the two different data sets; the embeddings appear visually consistent.
We also note that both $\mathcal{Y}_1$ and $\mathcal{Y}_2$ contain points that are projections of $\mathbf{Y}(1500), \dots, \mathbf{Y}(2000)$.
We therefore compute the correlation between the embedding coordinates for these points common to $\mathcal{Y}_1$ and $\mathcal{Y}_2$.
We obtain a correlation of 0.97 and 0.95 for the first and second NIV, respectively, indicating that the two embeddings are in quantitative
agreement with each other.
We would like to note that both $\mathcal{Y}_1$ and $\mathcal{Y}_2$ are sufficiently high-dimensional (``rich enough'') to allows us to recover
the common underlying two-dimensional manifold.

We then use NIV together with Laplacian Pyramids to estimate the values of $S:E$ and $F:E^{*}$ for $\mathcal{Y}_2$.
Because $\mathcal{Y}_1$ and $\mathcal{Y}_2$ are measured for different components, there is no simple way to estimate $S:E$ and $F:E^{*}$ directly in the observation space.
Instead, we must first embed the data into the NIV space so that we can compute neighbors {\em between the two data sets}.
We use $\mathcal{Y}_1$ to train an LP function from the two-dimensional NIV embedding to $S:E$ and $F:E^{*}$.
We then use this function to predict the values  of $S:E$ and $F:E^{*}$ for $\mathcal{Y}_2$, using the computed NIV embedding for $\mathcal{Y}_2$.
In this way, we are exploiting the fact that the NIV embedding is intrinsic and consistent between the two data sets, even though the two data sets contain measurements of different chemical species.

\begin{figure*}[ht]
    \includegraphics[width=6in]{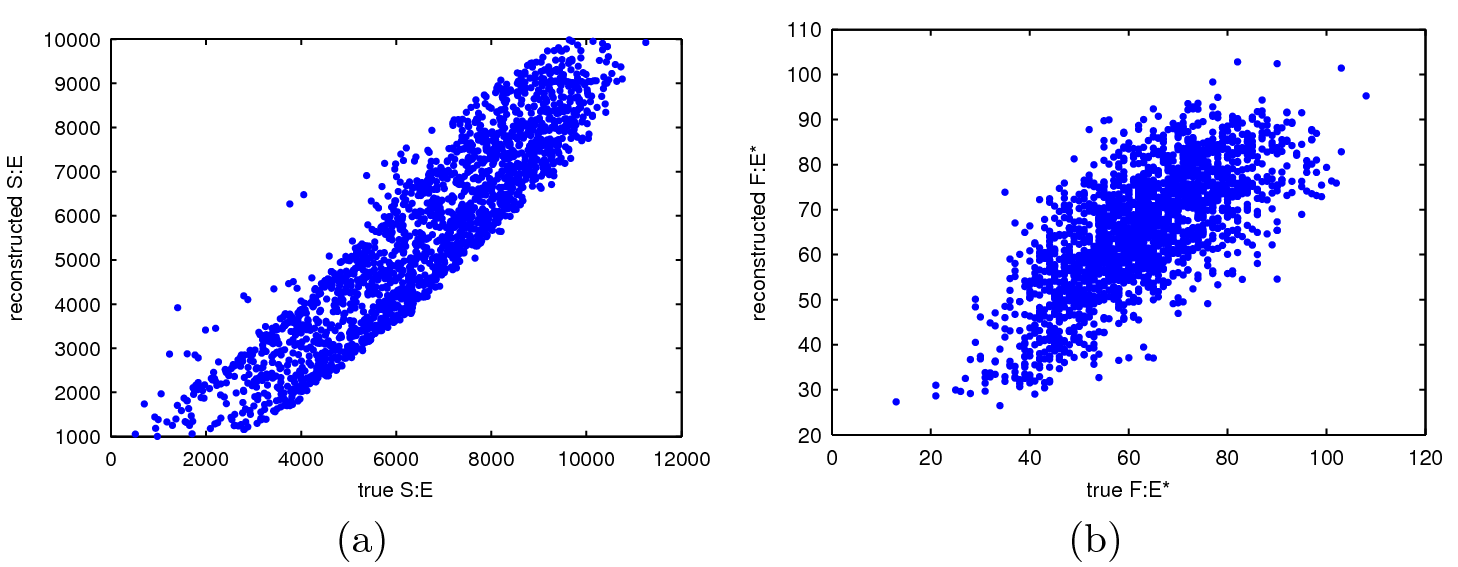}
    \caption{LP reconstructions of (a) $S:E$ and (b) $F:E^{*}$ for $\mathcal{Y}_2$, using $\mathcal{Y}_1$ as training data.}
    \label{fig:rxn_recon}
\end{figure*}

The results of the LP prediction are shown in Figure \ref{fig:rxn_recon}.
The normalized mean-squared errors between the true and estimated values for $S:E$ and $F:E^{*}$, defined as $\frac{\langle (y_{true}-y_{pred})^2 \rangle}{\langle y_{true}^2 \rangle}$, are 0.0372 and 0.0287 , respectively.
Therefore, we can effectively estimate the unobserved components in the reaction network using NIV together with LP.

\subsection{Alanine Dipeptide}

\begin{figure*}[ht]
    \includegraphics[width=6in]{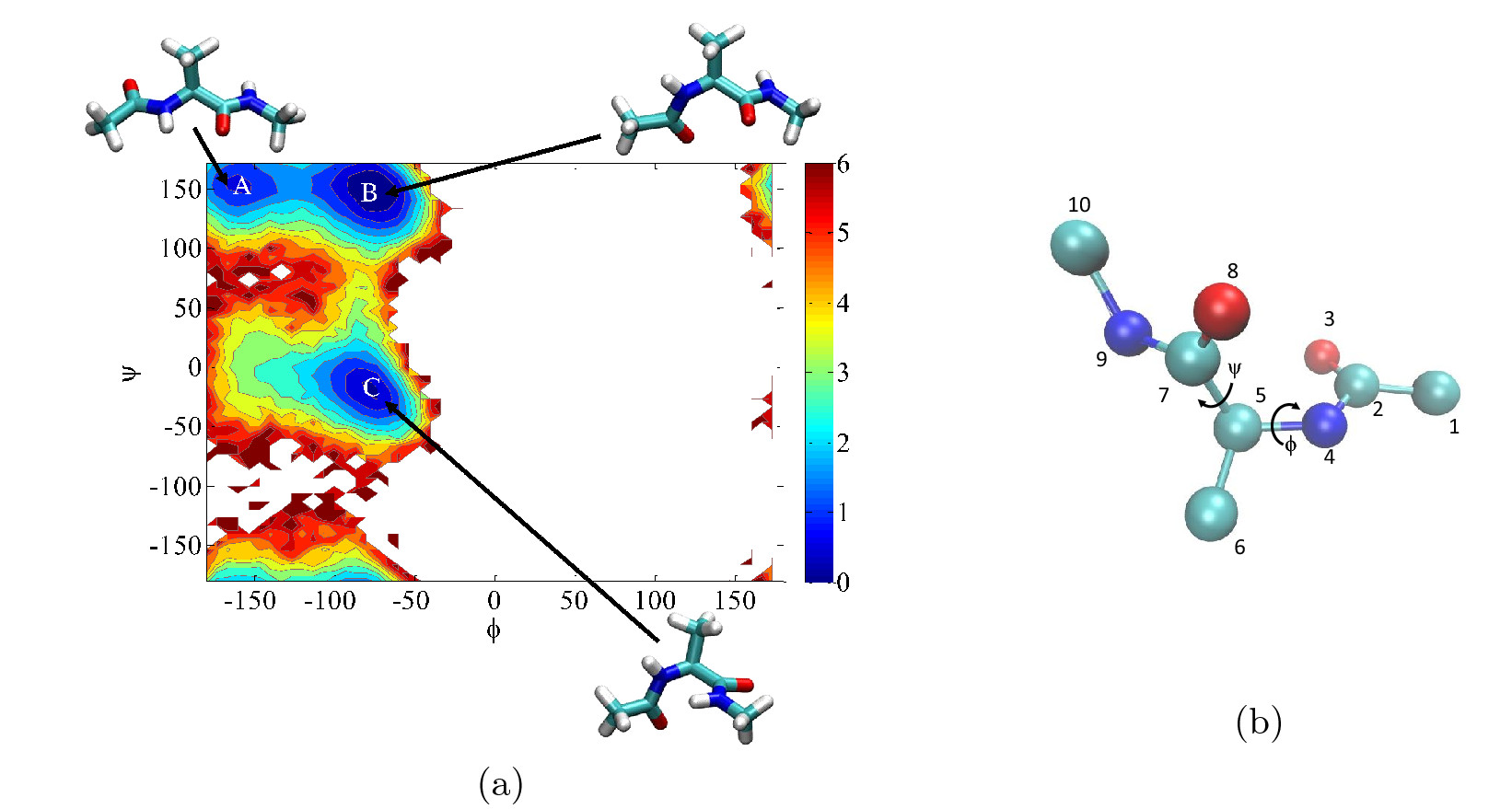}
    \caption{(a) Free energy surface for Ala2. The relevant minima are labeled A, B, and C, and the corresponding molecular configurations are shown.
    (b) Sample representative molecular structure of Ala2, excluding the hydrogens. The atoms are numbered and the two dihedral angles $\phi$ and $\psi$ are indicated.}
    \label{fig:ala_fes}
\end{figure*}

Our second example comes from the molecular dynamics simulation of a small peptide fragment.
Alanine dipeptide (Ala2) is often used as a ``prototypical'' protein caricature for simulation studies
\cite{apostolakis1999calculation, bolhuis2000reaction, chekmarev2004long, ma2005automatic, frewen2009exploration, ferguson2011integrating}.
We simulate the motion of Ala2 in explicit solvent using the AMBER 10 molecular simulation package \cite{case2008Amber} with an
optimized version \cite{best2009optimized} of the AMBER ff03 force field \cite{duan2003point}.
The molecule is solvated with 638 TIP3P water molecules \cite{jorgensen1983comparison}
with periodic boundary conditions, and the particle mesh Ewald method is used for long-range electrostatic interactions \cite{essmann1995smooth}.
The simulation is performed at constant volume and temperature (NVT ensemble), with the temperature being maintained at 300~K with a Langevin thermostat \cite{loncharich1992langevin}.
Hydrogen bond lengths are fixed using the SHAKE algorithm \cite{ryckaert1977numerical}.
The two dihedral angles $\phi$ and $\psi$ are known to parameterize the free energy surface, which contains three important minima (labeled A, B, and C, see Figure \ref{fig:ala_fes}).
Our simulations are concentrated around minimum B in the free energy surface, located at $\phi \approx -65^{\circ}$, $\psi \approx 150^{\circ}$.
We start many simulations at $10^{\circ}$ away from the minimum, and allow the simulations to each run for 0.1~ps, while recording the configuration of Ala2 every 1~fs (therefore, each trajectory is 100 points long).
Configurations are recorded with all atoms except the hydrogens.

We first compare NIV with direct diffusion maps \cite{coifman2005geometric}, an established nonlinear dimensionality reduction technique.
We consider 10,000 data points from our simulation $\mathbf{Y}(1), \dots, \mathbf{Y}(10000) \in \mathbb{R}^{30}$; every 100 data points comes from a continuous simulation trajectory.
We construct two data sets:
$\mathcal{Y}_{even} = \{\mathbf{Y}(t) \text{ restricted to atoms 2, 4, 6, 8, and 10}: t=1, \dots, 10000 \}$,
and $\mathcal{Y}_{odd} = \{\mathbf{Y}(t) \text{ restricted to atoms 5, 7, and 9}: t=1, \dots, 10000 \}$ (see Figure \ref{fig:ala_fes} for the atom indexing).
We then compute the NIV and diffusion maps embeddings for $\mathcal{Y}_{even}$ and $\mathcal{Y}_{odd}$;
for NIV, we compute the covariances as in \eqref{eq:cov}, with $L=10$.

\begin{figure*}[ht]
    \includegraphics[width=6in]{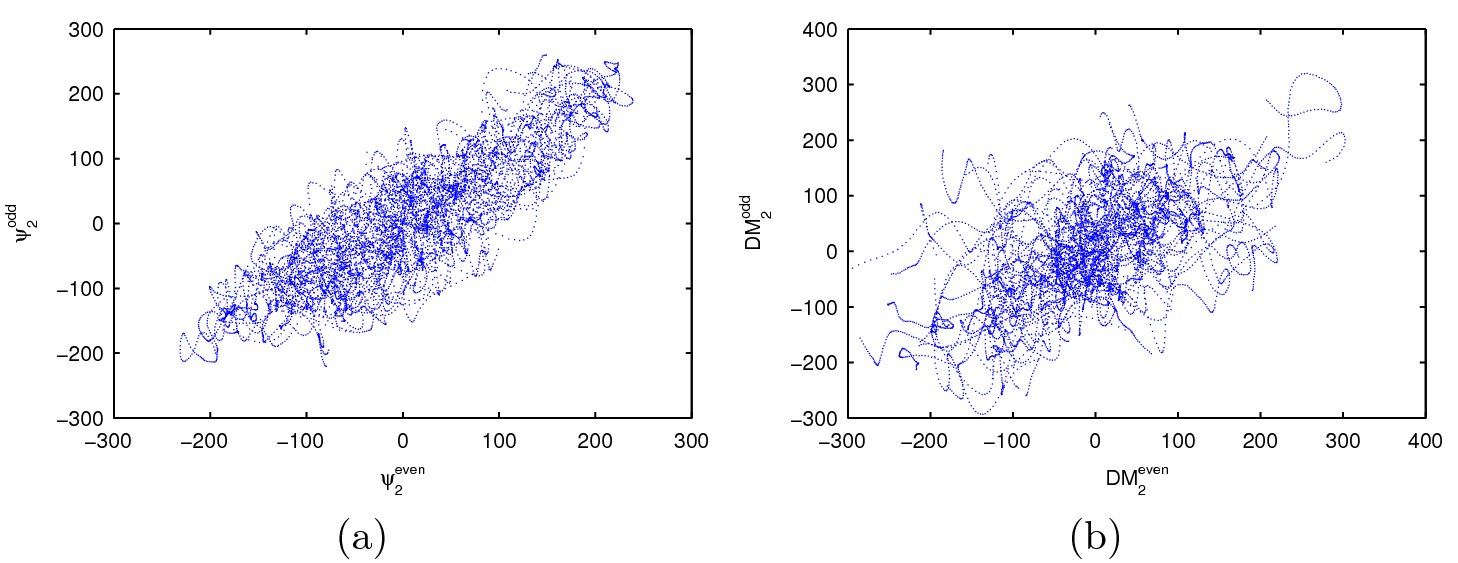}
    \caption{(a) Correlation between the second NIV computed using the atoms 2, 4, 6, 8, and 10 ($\psi_2^{even}$) and the second NIV computed using atoms 5, 7, and 9 ($\psi_2^{odd}$). 
    (b) Correlation between the second DM computed using the atoms 2, 4, 6, 8, and 10 ($DM_2^{even}$) and the second DM computed using atoms 5, 7, and 9 ($DM_2^{odd}$).
    The correlations for the first (not shown) and second NIV coordinates are found to be 0.62 and 0.84, respectively.
    The correlations for the first (not shown) and second DM coordinates are found to be 0.54 and 0.60, respectively. }
    \label{fig:ala_corr}
\end{figure*}

The correlation between the NIV coordinates for the two data sets and the diffusion map (DM) coordinates for the two data sets are shown in Figure \ref{fig:ala_corr}.
The correlation between the two NIV embeddings is higher than the correlation between the two diffusion map embeddings.
Therefore, it appears advantageous to use NIV over diffusion maps if one wishes to obtain a consistent embedding and merge data sets from different observation domains (as long as the two main assumptions underpinning the NIV algorithm hold).

We then use NIV together with LP to predict the conformation of Ala2 when we only observe some of the atoms.
We have 20000 data points $\mathbf{Y}(1), \dots, \mathbf{Y}(20000)$, where every 100 data points come from one continuous simulation trajectory.
Our first data set (which will serve as our training data for LP), $\mathcal{Y}_{all}$,
consists of the first 10000 data points ($\mathcal{Y}_{all} = \{\mathbf{Y}(t): t=1, \dots, 10000\}$).
Our second data set (which will serve as our test data), $\mathcal{Y}_{odd}$, consists of the last 12000 data points restricted to {\em only} the odd atoms
($\mathcal{Y}_{odd} = \{ \mathbf{Y}(t) \text{ restricted to the odd atoms}: t = 8001, \dots, 20000\}$).
We compute the covariances as in \eqref{eq:cov} with $L=15$.
%
We compute the NIV embedding for the training data $\mathcal{Y}_{all}$ and the test data $\mathcal{Y}_{odd}$; we then use LP interpolation from the training data to predict the location of all the atoms for each point in the test data.

\begin{figure*}[ht]
    \includegraphics[width=6in]{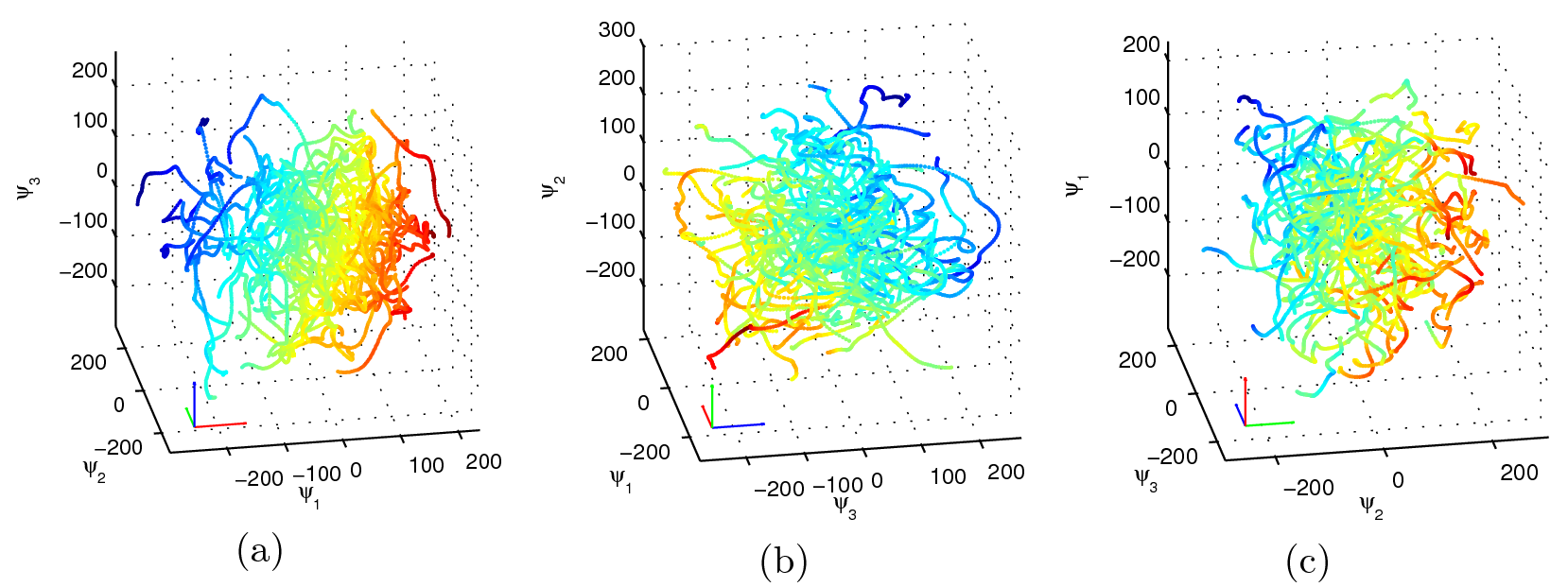}
    \caption{The 3-dimensional NIV embedding for Ala2 computed using $\mathcal{Y}_{all}$, colored by (a) the y-coordinate of the first atom, (b) the dihedral angle $\phi$, and (c) the dihedral angle $\psi$. Each embedding is rotated so that the correlation between the colors and the relevant NIV can easily be seen.}
    \label{fig:ala_embed}
\end{figure*}

The NIV embedding for the training data $\mathcal{Y}_{all}$ is shown in Figure \ref{fig:ala_embed}.
The embedding is three-dimensional, and visual inspection reveals that each coordinate can be directly linked with one physical variable:
the first coordinate describes the flipping of atoms 1 and 3, the second coordinate describes the dihedral angle $\phi$, and the third coordinate describes the dihedral angle $\psi$.
We calculate the correlation between the embedding coordinates for the points in $\mathcal{Y}_{all}$ and $\mathcal{Y}_{odd}$
that come from the common simulation data points $\mathbf{Y}(8001), \dots, \mathbf{Y}(10000)$.
The embeddings for the two data sets are found to be fairly consistent, with correlations of 0.97, 0.72, 0.85 for the first, second, and third NIV, respectively.


\begin{figure*}[ht]
  \centering
        \includegraphics[width=4.5in]{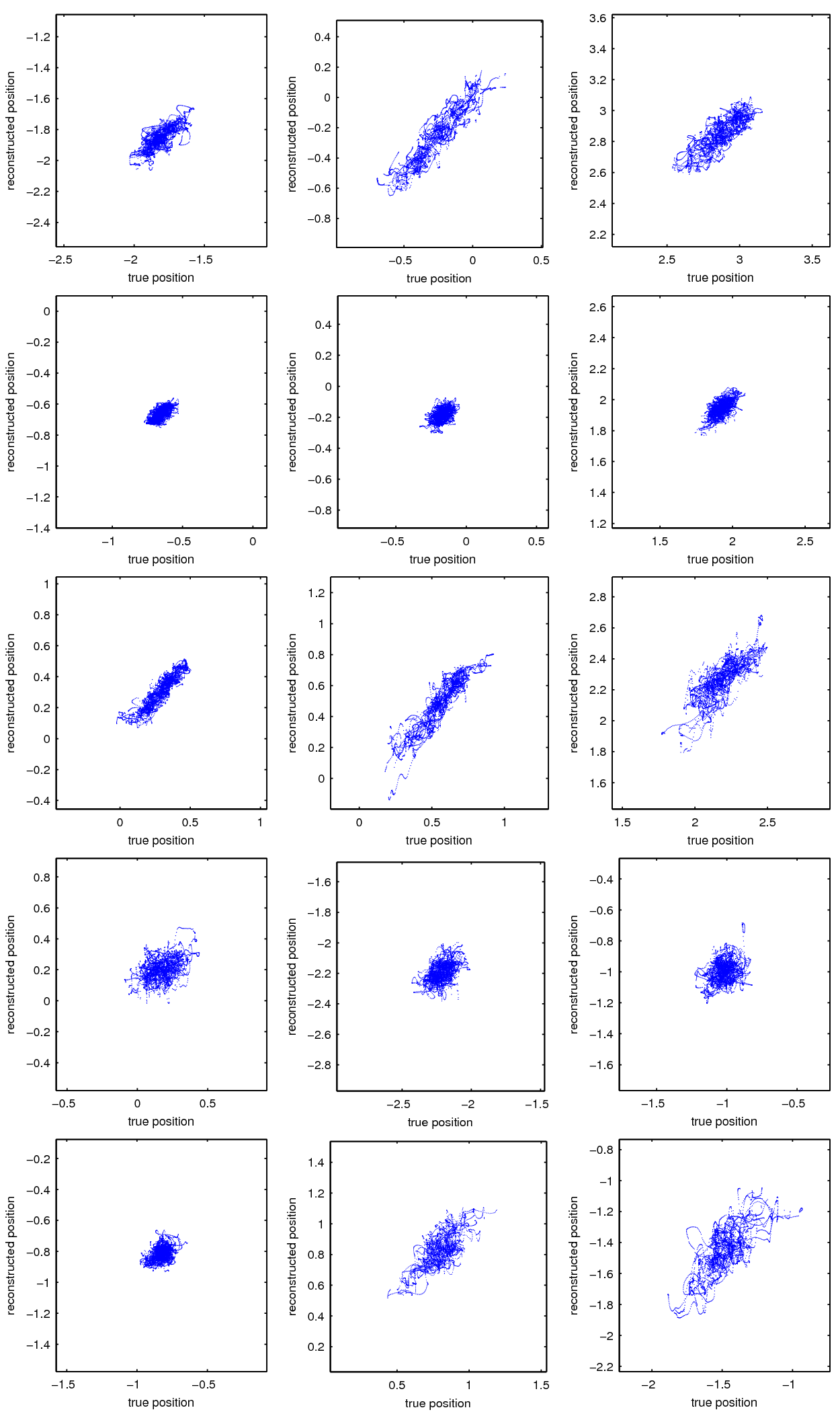}
  \caption{The correlation between the true position and the reconstructed position (using LP) for the test data. The columns correspond the x-, y-, and z-coordinates, and the rows correspond to atoms 1, 2, 3, 6, and 8.}
  \label{fig:ala_recon}
\end{figure*}

\begin{figure*}[ht]
    \centering
    \includegraphics[width=4in]{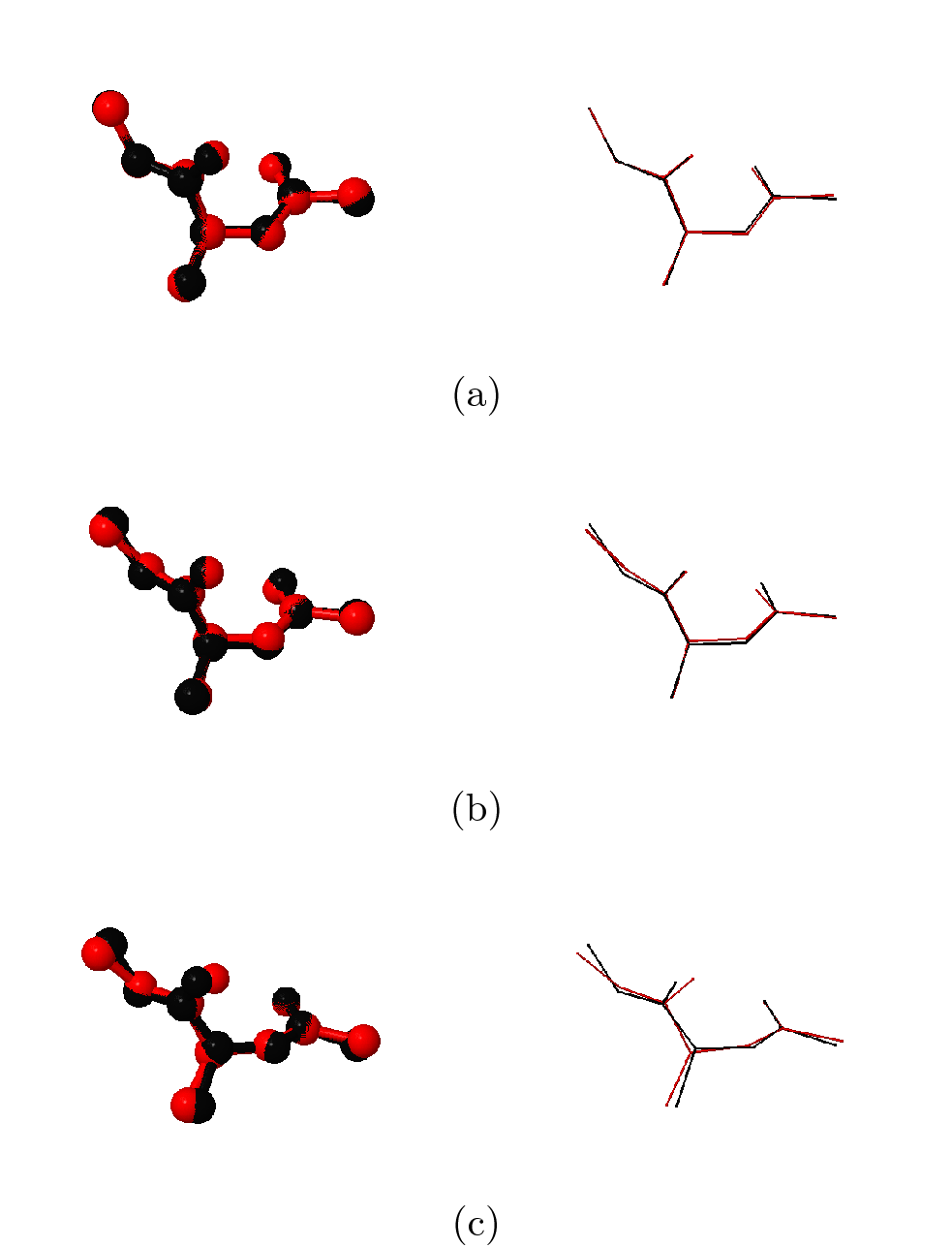}
    \caption{True structure (black) and reconstructed structure (red) for three different data points. Each data point is shown in ``ball-and-stick'' representation (left) and ``wireframe'' representation (right) so that the discrepancies can easily be seen between the different configurations.}
    \label{fig:ala_molecules}
\end{figure*}

Figure \ref{fig:ala_recon} shows the reconstructed position from partial observation versus true position for certain selected atoms.
The strong correlation between the true and reconstructed positions is easier to appreciate for
atoms that move substantially within the data set (such as atoms 1 and 3).
Figure \ref{fig:ala_molecules} shows molecular structures for the true and reconstructed configurations for selected data points;
there is qualitative agreement between the true and reconstructed configurations.

\begin{figure*}
    \includegraphics[width=3in]{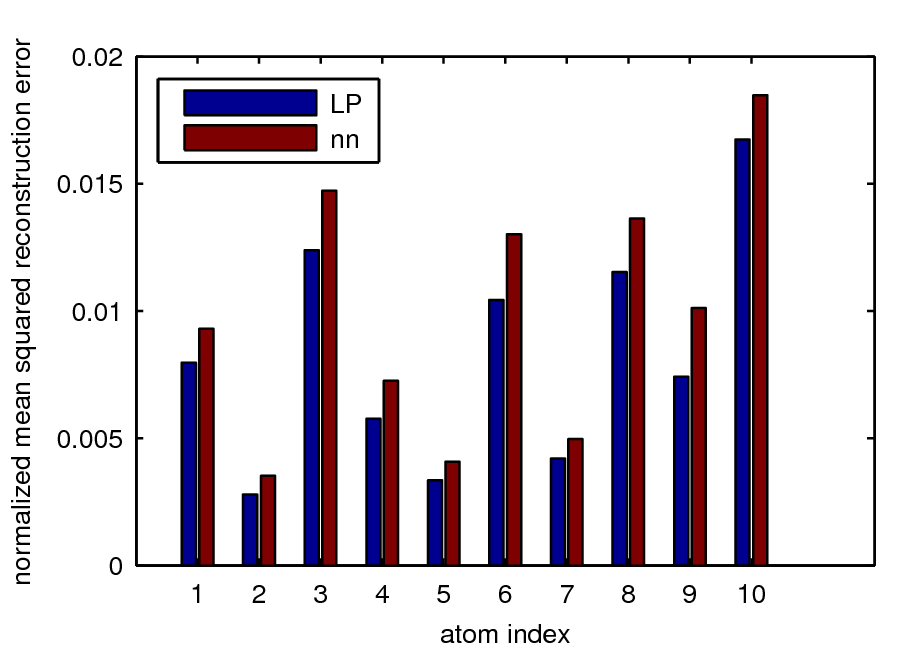}
    \caption{Mean squared error of reconstructed position, normalized by the average bond length, for each atom in Ala2.
    The positions were reconstructed using both LP and nearest neighbor (nn) interpolation .}
    \label{fig:ala_mse}
\end{figure*}

For a brief comparison of LP over other reconstruction techniques, we also reconstruct configurations from the
NIV components using simple nearest-neighbor interpolation.
The average reconstruction error, scaled by the average bond length within the molecule, is shown in Figure \ref{fig:ala_mse};
LP arguably outperforms simple nearest neighbor search for all of the atoms.

\section{Conclusions} \label{sec:conclusions}
We have used Nonlinear Intrinsic Variables to analyze two complex atomistic simulations: a stochastic simulation of a chemical reaction network and a molecular dynamics simulation of alanine dipeptide.
In both examples, we were able to uncover the intrinsic variables governing the underlying stochastic process, which are independent
of the particular measurement or observation  of the system (under the conditions mentioned).
The uniqueness of the embedding coordinates allowed us to compare and merge data sets from different measurement functions,
and therefore allowed us to use an interpolation/extension scheme (here Laplacian Pyramids) to complete partial observations.
Different interpolation techniques (e.g. kriging \cite{matheron1963principles, matheron1973intrinsic}, geometric harmonics \cite{coifman2006geometric},  versions of the Nystr\"{o}m extension) can and should
be explored, since the performance of such techniques may well be problem dependent, especially for multiscale, complex simulation data.

There are many open questions leading to interesting research directions to be explored.
In this work, we considered data sets that consist of different partial observations, but in which each data set samples the entire underlying manifold in
what we loosely referred to as a ``representative enough" way.
However, NIV could also be used to merge data sets when each data set samples only a portion of the manifold, provided there is enough overlap to ``register'' the embeddings.
Merging data sets that come from different portions of the manifold would not only require scaling the embedding coordinates,
but also shifting and possibly permuting the embedding coordinates (in this spirit, see the discussion in Lafon {\em et al.} \cite{lafon2006data}).
The ability to merge data from different regions would then allow us to analyze systems where complete sampling is computationally intractable,
such as molecular systems with several high energy barriers separating regions of state space.

%
%
Other issues, such as accurately estimating the covariance matrices required for the computation of the Mahalanobis distance, are also of current research interest.
It is clearly necessary to link this type of calculation with modern estimation techniques for (multiscale) diffusions \cite{ait2002maximum, ait2003effects, ait2008closed}
to test the appropriateness of the window sampling lengths selected; this will determine the accuracy of the noise covariance estimation by eliminating
the bias due to drift variations.
We are confident that the exploration of these open questions will enable the use of our methodology in many  interesting applications,
such as merging data from molecular simulations at different levels of granularity, or merging simulation data with experimental observations.

\begin{acknowledgments}
C. J. D. would like to acknowledge support from the US Department of Energy Computation Science Graduate Fellowship, grant number DE-FG02-97ER25308.
I. G. K. would like to acknowledge support from the US Department of Energy, grant numbers DE-FG02-10ER26024 and DE-FG02-09ER25877.
\end{acknowledgments}

%

\appendix

\section{Chemical Reaction Network Parameters} \label{app:rxn}

We consider the following network of chemical reactions.
\begin{equation}
\begin{array}{rcl}
E + S \overset{e_1}{\underset{e_{-1}}{\leftrightharpoons}} & E:S & \overset{e_2}{\rightarrow} E + S^{*} \\
S + E \overset{b_1}{\underset{b_{-1}}{\leftrightharpoons}} & S:E & \overset{b_2}{\rightarrow} S + E^{*}\\
D + S^{*} \overset{d_1}{\underset{d_{-1}}{\leftrightharpoons}} & D:S^{*} & \overset{d_2}{\rightarrow} D + S\\
F + E^{*} \overset{f_1}{\underset{f_{-1}}{\leftrightharpoons}} & F:E^{*} & \overset{f_2}{\rightarrow} F + E
\end{array}
\end{equation}

\begin{widetext}
In the limit of a large number of molecules, the dynamics of this network is governed by the following ODEs.
\begin{equation}
    \begin{array}{rcl}
        S^\prime & = & -e_1 S E + e_{-1} E:S -b_1 S E + b_{-1} S:E + b_2 S:E + d_2 D:S^{*} \\
        E^\prime & = & -e_1 S E + e_{-1} E:S + e_2 E:S - b_1 S E +b_{-1} S:E + f_2 F:E^{*} \\
        E:S^\prime & = & e_1 S E -e_{-1} E:S -e_2 E:S \\
        S:E^\prime & = & b_1 S E -b_{-1} S:E -b_2 S:E \\
        {S^{*}}^\prime & = & e_2 E:S -d_1 D S^{*} + d_{-1} D:S^{*} \\
        {E^{*}}^\prime & = & b_2 S:E -f_1 F E^{*} + d_{-1} F:E^{*} \\
        D^\prime & = & -d_1 D S^{*} + d_{-1} D:S^{*} + d_2 D:S^{*} \\
        F^\prime & = & -f_1 F E^{*} + f_{-1} F:E^{*} + f_2 F:E^{*} \\
        {D:S^{*}}^\prime & = & d_1 D S^{*} - d_{-1} D:S^{*} -d_2 D:S^{*} \\
        {F:E^{*}}^\prime & = & f_1 FE^{*} - f_{-1} F:E^{*} -f_2 F:E^{*}
    \end{array}
\end{equation}
\end{widetext}

We can write four balance equations for the conservation of total $S$, $E$, $D$, and $F$.
\begin{equation}
    \begin{array}{rcl}
        S_T & = & S^{*} + S + E:S + S:E + D:S^{*} \\
        E_T & = & E^{*} + E + E:S + S:E + F:E^{*} \\
        D_T & = & D + D:S^{*} \\
        F_T & = & F + F:E^{*}
    \end{array}
\end{equation}

\begin{widetext}
We choose to eliminate $S^{*}$, $E^{*}$, $D$, and $F$ from the system of ODEs.
We therefore obtain a system of 6 ODEs.
\begin{equation}
    \begin{array}{rcl}
        S^\prime & = & -e_1 S E + e_{-1} E:S -b_1 S E \\
                        &   & + b_{-1} S:E + b_2 S:E + d_2 D:S^{*} \\
        E^\prime & = & -e_1 S E + e_{-1} E:S + e_2 E:S \\
                        &   & - b_1 S E +b_{-1} S:E + f_2 F:E^{*} \\
        E:S^\prime  & = & e_1 S E -e_{-1} E:S -e_2 E:S \\
        S:E^\prime & = & b_1 S E -b_{-1} S:E -b_2 S:E \\
        {D:S^{*}}^\prime & = & d_1 (D_T - D:S^{*}) (S_T - S - E:S - S:E - D:S^{*}) - d_{-1} D:S^{*} -d_2 D:S^{*} \\
        {F:E^{*}}^\prime & = & f_1 (F_T - F:E^{*}) (E_T - E - E:S - S:E - F:E^{*}) - f_{-1} F:E^{*} -f_2 F:E^{*}
    \end{array}
\end{equation}
\end{widetext}

Alternatively, we can write the rates for the 12 chemical reactions as
\begin{equation}
    \begin{array}{rcl}
        r_1 & = & e_1 S E \\
        r_2 & = & e_{-1} E:S \\
        r_3 & = & e_2 E:S \\
        r_4 & = & b_1 S E \\
        r_5 & = & b_{-1} S:E \\
        r_6 & = & b_2 S:E \\
        r_7 & = & d_1 (D_T - D:S^{*}) (S_T - S - E:S - S:E - D:S^{*}) \\
        r_8 & = & d_{-1} D:S^{*} \\
        r_9 & = & d_2 D:S^{*} \\
        r_{10} & = & f_1 (F_T - F:E^{*}) (E_T - E - E:S - S:E - F:E^{*}) \\
        r_{11} & = & f_{-1} F:E^{*} \\
        r_{12} & = & f_2 F:E^{*}
    \end{array}
\end{equation}

For the Gillespie SSA, we use these rates to adjust the {\em number} of each molecule, depending on which reaction occurs.
We take the volume of the reactor $V=10^5$.
We use the parameters $b_1=5/V$, $d_1=0.0009/V$, $e_1=0.1/V$, $f_1=0.1/V$, $b_{-1} = 10.6$, $d_{-1}=0.05$, $e_{-1}=0.5$, $f_{-1} =0.01$, $b_2=0.4$, $d_2=0.85$, $e_2=0.05$, and $f_2=2$.
We take $S_T=E_T=D_T=1V$, and $F_T=0.02V$, where $S_T$, $E_T$, $D_T$, and $F_T$ are total number of $S$, $E$, $D$, and $F$, respectively.
In this parameter regime, the relevant timescales around the steady state ($-1/\lambda_i$, where $\lambda_i$ are the eigenvalues of the Hessian) are 1176, 9.731, 1.594, 1.111, 0.4975, 0.06498.
Therefore, we choose to evolve forward for 10 time units to find points on a perceived two-dimensional manifold.

\end{document}